\documentstyle[12pt,epsfig]{article} 
\def\baselinestretch{1.3}
\newcommand{\ba}{\begin{array}}
\newcommand{\ea}{\end{array}}
\newcommand{\bd}{\begin{displaymath}}
\newcommand{\ed}{\end{displaymath}}
\newcommand{\be}{\begin{equation}}
\newcommand{\ee}{\end{equation}}
\newcommand{\bea}{\begin{eqnarray}}
\newcommand{\eea}{\end{eqnarray}}

\def\q2 {q^2}

\def\miss {\hspace{-0.5cm}\slash~~}

\def\rslep {\tilde{e_R}}
\def\rsnu {\tilde{\nu}_R}
\def\snu {\tilde{\nu}}
\def\lslep {\tilde{e_L}}
\def\stau {\tilde{\tau}}
\def\mer {m_{\rslep}}
\def\mmr {m_{\tilde{\mu}_R}}
\def\mml {m_{\tilde{\mu}_L}}
\def\mel {m_{\lslep}}

\def\bt{\begin{table}}
\def\et{\end{table}}

\catcode`@=11 
\def \gsim{\mathrel{\mathpalette\@versim>}}
\def \lsim{\mathrel{\mathpalette\@versim<}}
\def \@versim#1#2{\lower0.4ex\vbox{\baselineskip\z@skip\lineskip\z@skip
     \lineskiplimit\z@\ialign{$\m@th#1\hfil##\hfil$%
     \crcr#2\crcr\sim\crcr}}}
\catcode`@=12 

\begin{document}
\begin{flushright}
{\large  HRI-P-07-01-001}
\end{flushright}

\begin{center}
{\Large\bf Right-chiral sneutrinos and long-lived staus:
event characteristics at the Large Hadron Collider }\\[20mm]
Sudhir Kumar Gupta \footnote{E-mail: guptask@mri.ernet.in}, 
Biswarup Mukhopadhyaya\footnote{E-mail: biswarup@mri.ernet.in}
and Santosh Kumar Rai \footnote{E-mail: skrai@mri.ernet.in} \\
{\em Harish-Chandra Research Institute,\\
Chhatnag Road, Jhusi, Allahabad - 211 019, India}, 
\\[20mm] 
\end{center}
\begin{abstract}
We investigate the signals of a supersymmetric scenario
where the lighter tau-sneutrino is the lightest supersymmetric particle,
while the lighter stau-state is the next lightest. We confirm 
that such a scenario can be motivated within the framework of
minimal supergravity, with just the addition of a right-chiral 
neutrino superfield. Such a spectrum leads to rather unusual
signals of supersymmetry, showing stable tracks
of the stau in the muon chambers. We have studied two types
of signals, namely, (a) two or more hard jets and two stau tracks,
and (b) hard jets with two muons and two stau tracks. We demonstrate
that the stau tracks can be distinguished from the muonic ones
through proper kinematic cuts which also enable one to remove
all standard model backgrounds.
\end{abstract}

\vskip 1 true cm

\newpage
\setcounter{footnote}{0}

\def\baselinestretch{1.5}

\section{Introduction}

Supersymmetry (SUSY) is still the most sought-after new physics option
around the TeV scale, and concerted efforts are on to discover it in
any of its possible ramifications at the Large Hadron Collider (LHC).
Apart from stabilizing the electroweak symmetry breaking sector and
providing a rather tantalizing hint of Grand Unification, SUSY (in the
R-parity conserving version) also provides a cold dark matter candidate
in the form of the stable lightest supersymmetric particle (LSP). The
most common practice is to assume that the lightest neutralino is the
LSP \cite{reviews1}, and the `standard' signals such as 
(jets + ${E_T}\miss$),
(dileptons + ${E_T}\miss$) etc. are widely studied under this
assumption \cite{reviews2}. After decades of rather exhaustive 
investigation, two
questions that are now being asked are, (a) how tractable are SUSY
signals with other types of LSP? And (b) is it possible to have
quasi-stable but cosmologically allowed charged particles in a SUSY
scenario, with distinct collider signatures? As we shall see below,
the answers to both questions can be of diverse nature but nonetheless
related, and are worth exploring in detail before the LHC takes off.

It must be realized at the same time that hints of new physics stare us
in the face at a scale much lower than a TeV, as one finds
incontrovertible evidence of neutrino masses and mixing of rather
unexpected nature \cite{revnu}. 
The shortest step beyond the standard electroweak
model, which provides an explanation of them, consists in hypothesizing
a right-handed neutrino for each fermion family \cite{rneut}. 
If one accepts such a
hypothesis, and at the same time depends on SUSY for the stabilization
of the electroweak scale, then three additional members are added to
the assortment of R-odd (super)particles, namely, the right-chiral
sneutrinos. Interestingly, such sneutrinos are perfectly inactive with
respect to all the gauge interactions, and have only couplings
proportional to the neutrino masses. Of course, the physical states in
each family involve a tiny mixing of both chiral types. The
non-interactive nature of the dominantly right-chiral state relative to
the left-chiral one makes it possible for the former (as opposed to the
latter \cite{hebbeker}) to be a dark matter candidate \cite{arkani}. 
Therefore, it makes perfect
sense to envision a SUSY spectrum with such a sneutrino as the LSP.
 
In this paper, we study some signals of such a dominantly
right-sneutrino LSP (${\snu}_1$) at the LHC. When this happens, a
number of candidates exist for the next-to-lightest SUSY particle
(NLSP).  These include the dominantly left-sneutrino state
(${\snu}_2$), a slepton (especially the stau, with its opportunity
to have a rather light mass eigenstate with large $\tan \beta$, $\tan
\beta$ being the ratio of the vacuum expectation values of the two
Higgs doublets), a neutralino, a chargino, or even a stop in special
cases. Among these, we consider the particular case of a stau NLSP
because the collider signals are rather typical in the pronounced
difference with the neutralino LSP scenario. The consequences of a
spectrum with, for example, a stop NLSP has been studied in the
literature \cite{gopala}.

Interestingly, for a ${\snu}_1$ LSP, any NLSP will decay into it
only through an interaction strength proportional to the neutrino mass.
This is because Yukawa interactions will invariably involve this mass,
while gauge interaction will depend on the admixture of the $SU(2)_L$
doublet component in ${\snu}_1$ proportional to the neutrino
Yukawa coupling. Thus the decay of the NLSP to the LSP is too
suppressed over most of the viable parameter space to take place within
the detector. While the LSP still contributes (at least partially) to
cold dark matter, the NLSP appears to be stable, as far as collider
detectors are concerned. Consequently, SUSY signals in the case
considered by us will be marked by charged tracks characteristic of
staus.

`Stable stau' phenomenology of the kind mentioned above can in general
be of several origins, namely,

\begin{itemize}
\item Gravitino LSP in a supergravity (SUGRA) theory \cite{gravitinoLSP}. 
Here the
gravitino is of such mass as to be a cold dark matter candidate, and
its coupling to the stau (being inversely proportional to the gravitino
mass) is suppressed enough to prevent the latter from decaying within
the detector.

\item Stau-NLSP in gauge mediated SUSY breaking (GMSB) models
\cite{gmsbnlsp}. Here the
stau may be stable if the effective F-term leading to the dynamical
generation of gravitino mass is on the higher side (above about $(10^7
GeV)^2$).

\item A neutralino LSP nearly degenerate with a stau NLSP \cite{coanni}. 
This is the so-called co-annihilation region, with the small mass difference
suppressing the decay of the NLSP.

\item An LSP dominated by the right-chiral sneutrino \cite{Moroi2}, 
with a stau NLSP.
As we shall see, such a possibility is quite viable in a SUGRA scenario
admitting a right-neutrino superfield in the observable sector.
\end{itemize}

The present study concerns the last one of the above scenarios.
Although long-lived stau NLSP's arising from one or the other of the
possibilities mentioned above are frequently discussed nowadays, our
purpose is to specially emphasize the following points:

(a) The stau can become an NLSP, and the corresponding right
sneutrino, the LSP, in rather natural regions of the SUGRA parameter
space, provided that right sneutrino mass is allowed to evolve from the
common scalar mass at high scale.

(b) The stau NLSP is likely to leave charged tracks on reaching the
muon detector \cite{longlived}. 
Ways of distinguishing it from a muon have been
suggested on the basis of the thickness of the track, and also through
the time delay between the inner tracking chamber and the muon
detector. Moreover, the use of stoppers has been sometimes advocated
for intercepting long-lived staus, whereby their late decays can be
studied in underground chambers at a later stage \cite{stoppers}. 
While the efficacy of
such methods is admitted, we wish to point out that the very kinematic
properties of the tracks in the muon chamber set the long-lived staus
apart from the muons in a conspicuous fashion, and the characteristic
signal events of such a scenario can be separated from the standard
model (SM) backgrounds in a rather straightforward manner.

In section 2 we discuss how the ${\snu}_1$-LSP, stau-NLSP
scenario can arise naturally within the SUGRA framework including a
right-chiral neutrino superfield for each generation. Section 3 
contains discussions on the two principal signals of such a scenario,
which can originate in the production and cascade decays of strongly
interacting superparticles. These are the (hard jets +
$\stau$-pair) and (hard jets + $\stau$-pair + dimuon)
signals, respectively.\footnote{Although di-electron signals are as
important as dimuons, we demonstrate our predictions with reference to
the latter only, since the stable staus may appear like muons. Hence
the challenge of separating such `fake' muons with real ones is best
reflected in a analysis involving dimuon samples.}. We summarize and
conclude in section 4.

\section{Right sneutrino LSP in supergravity}
The simplest extension to the SM spectrum, which would explain
non-vanishing neutrino masses \cite{neuexp}, 
is to add one right-handed neutrino for each generation.
However, having such small Dirac masses for the neutrinos would imply
that the neutrino Yukawa couplings are quite small. Considering
the different neutrino mass hierarchies as benchmark scenarios,
a normal (or inverted) hierarchy would imply that such couplings 
are at most of the order of $\sim 10^{-13}$, whereas they can be larger
by an order or so for the case of degenerate neutrinos.

Here we consider a scenario where lepton number is conserved. 
Thus the superpotential of the minimal SUSY standard model (MSSM)
is extended by just one term which, for a particular family,
is of the form
\bea
W_\nu^R=y_\nu \hat{H}_u \hat{L}\hat{\nu}^c_R
\eea
where $y_\nu$ is the Yukawa coupling, $\hat{L}$ is the left-handed lepton
superfield and $\hat{H}_u$ is the Higgs superfield responsible for giving 
mass to the $T_3=+1/2$ fermions. The above term in the superpotential 
obviously implies the inclusion of
right-handed sneutrinos in the particle spectrum. As discussed in the
introduction, such sneutrinos will have all their 
interactions proportional to the corresponding neutrino masses. So
the dominantly right-handed eigenstate of the tau-sneutrino might become a
possible candidate for the LSP. Working in the 
framework of minimal supergravity (mSUGRA) model of SUSY, we find 
that one can have the lightest sneutrino (composed mostly
of the right-chiral component)
as the LSP, consistent with all experimental bounds \cite{expbound} 
and also within the acceptable limits of dark matter density in the 
universe \cite{Moroi2,wmap}. 

The neutrinos obtain mass as:
\bea
m_\nu = y_\nu \left<H_u^0\right> = y_\nu v~sin\beta
\eea

The actual mass eigenvalues will of course depend on the Yukawa
couplings which may have a hierarchy among them, and will in general
be matrix-valued. However, the different neutrino mass hierarchies, 
and the consequent hierarchies in the Yukawa couplings,
do not affect the generic feature of the collider signal under consideration 
here, as the decay rate of the stau-NLSP is extremely suppressed in either of 
them. It is basically the smallness of the 
Yukawa coupling which plays a crucial role in our 
understanding of the spectrum and its consequent features.

Upon inclusion of right-chiral neutrino superfield into the SUGRA fold, 
the superparticle spectrum mimics the mSUGRA spectrum in all details except 
for the identity of the LSP.  
SUSY breaking effects are introduced explicitly as universal soft terms 
for scalars ($m_0$) and gauginos ($m_{1/2}$) 
together with the so-called A and B-parameters (of which the 
latter is determined by electroweak symmetry breaking conditions)
in the Lagrangian. Masses of squarks, sleptons and gauginos, all
the mass parameters in the Higgs sector as well as the Higgsino
mass parameter $\mu$ (upto a sign) are determined, once the
SUSY breaking parameters at a high scale (${\mathcal O}\sim10^{11}$ GeV) 
are specified. The mass terms for sneutrinos (neglecting inter-family 
mixing) arising in this manner are given by
\bea
-{\mathcal L}_{soft} \sim {M}_{{\rsnu}}^2 
|\widetilde{\nu}_R|^2 + (y_{\nu}{A}_\nu H_u.\widetilde{L} 
\widetilde{\nu}_R^c ~~ + ~~h.c.)
\eea
where $A_\nu$ is the term driving left-right mixing in the
scalar mass matrix, and is obtained by running of the
trilinear soft SUSY breaking term $A$.
The Yukawa couplings can cause large splitting in the third-family squark 
and sleptons masses while the first two families are effectively
degenerate. On the other hand, one expects minimal left-right mixing of
sneutrinos as the Yukawa couplings are all extremely small.

The mass-squared matrix for the sneutrino thus looks like
\bea
m_{\tilde{\nu}}^2 = \left ( \begin{array}{cc} {M}_{\tilde{L}}^2 +
\frac{1}{2}m_Z^2\cos 2\beta & y_\nu v({A}_\nu \sin\beta-\mu\cos\beta)\\
y_\nu v({A}_\nu \sin\beta-\mu \cos\beta) & 
{M}_{\tilde{\nu}_R}^2 \end{array} \right)
\eea
where ${M}_{\tilde{L}}$ is the soft scalar mass for the 
left-handed sleptons whereas the ${M}_{\tilde{\nu}_R}$ is 
that for the right-handed sneutrino. In general, 
 ${M}_{\tilde{L}} \ne {M}_{\tilde{\nu}_R}$ because of their different
evolution patterns as well as the D-term contribution for the former. 
While the evolution of all parameters of minimal SUSY
remain practically unaffected in this scenario, the  
right-chiral sneutrino mass parameter evolves \cite{arkani} at 
the one-loop level as: 
\bea
\frac{dM^2_{\rsnu}}{dt} = \frac{2}{16\pi^2}y^2_\nu~A^2_\nu ~~.
\eea

Clearly, the extremely small Yukawa couplings cause  ${M}_{\tilde{\nu}_R}$
to remain nearly frozen at the value $m_0$, whereas the other sfermion
masses are jacked up at the electroweak scale. Thus, for a wide range of
values of the gaugino mass, one naturally has sneutrino LSP's, which,
for every family,
is dominated by the right-chiral state:
\bea
\tilde{\nu}_1 = - \tilde{\nu}_L \sin\theta + \tilde{\nu}_R \cos\theta
\eea
The mixing angle $\theta$ is given as 
\bea
\tan 2\theta = \frac{2 y_\nu v\sin\beta |\cot\beta\mu -
A_\nu|}{m^2_{\tilde{\nu}_L}-m^2_{\rsnu}}
\eea
which is clearly suppressed due to $y_\nu$. It is to be noted
that all three (dominantly) right sneutrinos have similar fate
here, and one has near-degeneracy of three such LSP's. However,
of the three charged slepton families, the amount of left-right
mixing is always the largest in the third (being, of course,
more pronounced for large $\tan\beta$), and the lighter stau
(${\stau}_1$) often turns out to be the NLSP in such a scenario. 
\footnote{We have neglected inter-family mixing in the sneutrino 
sector in this study. While it is true that near-degenerate 
physical states makes their mixing a likely consequence, the
extent of such mixing is model-dependent, and does not generally
affect the fact that all cascades culminate in the lighter stau, 
so long as the latter in the NLSP, which is the scenario under study here.}

Thus the mSUGRA parameter set ($m_0,m_{1/2},A,sign(\mu)~{\rm and}~\tan\beta$)
in an R-parity conserving scenario can eminently lead to a spectrum where
all the three generation of right-sneutrinos will be either stable 
or metastable but very long-lived, and can lead to different decay chains
of supersymmetric particles, as compared to those with
a neutralino LSP. However, as we shall see below, the controlling
agent is the the lighter sneutrino eigenstate of the third family,
so long as the state ${\stau}_1$ is the lightest among
the charged sleptons.

\begin{figure}[ht]
\begin{center}
\epsfig{file=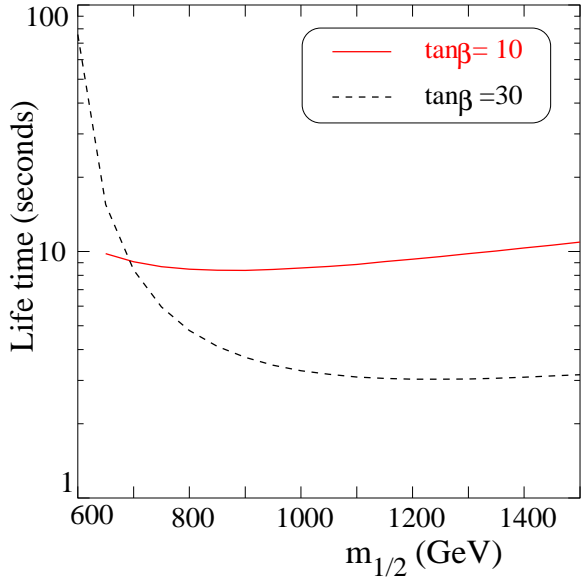,width=3.0in,height=3.0in}
\caption{\small
\it{Lifetime of stau NLSP against the
universal gaugino mass parameter $m_{1/2}$. Solid (red) line is
for $tan\beta=10$ and the dashed (black) line is for $tan\beta=30$.
Other SUGRA input parameters are: $m_0=100$ GeV, $A=100$ GeV,
$sgn(\mu)=1$.}}
\label{decay}
\end{center}
\end{figure}
We have chosen the above scenario to demonstrate our main point
for another reason. In general, apart from the stau, either a neutralino
or a chargino (being the remaining R-odd non-strongly interacting
particles) could be the NLSP. However, a neutralino NLSP will decay
into a neutrino and the sneutrino LSP, both of which are invisible. Thus
the collider signals will not be qualitatively different from 
those of a neutralino LSP model, irrespective of whether
the NLSP decays inside the detector or outside. A chargino NLSP, 
on the other hand, can make a difference in this respect through
charged tracks. However, it is difficult to construct a model
where the lighter chargino ($\chi^{\pm}_1$) 
is lighter than the lightest neutralino ($\chi^0_1$)
unless gaugino non-universality of a rather gross nature is
postulated. We have chosen to confine ourselves to a relatively
minimal high-scale picture, where such an option is somewhat
far-fetched. The remaining situation is one with a $\stau$ NLSP,
which is both (as we have shown above) easy to accommodate within
an mSUGRA scenario and gives qualitatively different SUSY signals.

Any (super)particle will have to decay into the lighter sneutrino of 
a particular family via either gauge interactions (such as
$\tilde{l}_L \longrightarrow W \tilde{\nu}_1$)  or Yukawa coupling
(such as $\tilde{l}_L \longrightarrow H^- \tilde{\nu}_1$ or
 $\tilde{\nu}_2 \longrightarrow h^0 \tilde{\nu}_1$). In the former case,
the decay depends entirely on the $\tilde{\nu}_L$ content
of  $\tilde{\nu}_1$, which again depends on the neutrino Yukawa coupling.
The same parameter explicitly controls the decay in the latter case, too.
Therefore, while the lighter sneutrinos of the first two families can in 
principle be produced from decays of the corresponding charged 
sleptons, such decays will be always suppressed compared to even 
three-body decays such as 
$\tilde{e}_1 (\tilde{\mu}_1) \longrightarrow e(\mu) \bar{\tau} {\stau}_1$.
For the $\stau$-NLSP, however, the only available decay channel
is $\stau_1 \longrightarrow W \tilde{\nu}_1$, which is driven
by the extremely small neutrino Yukawa coupling. 

This causes the NLSP to be a long-lived particle and opens
up a whole set of new possibilities for collider signatures for such
long-lived particles, while retaining contributions to cold
dark matter from the sneutrino LSP\footnote{In fact, 
the possibility of accounting at least partially for the cold
dark matter of the universe makes the right-sneutrino LSP scenario 
slightly more viable than one with a superlight gravitino playing
a similar role.}. The NLSP appears stable in collider
detectors and gives highly ionizing charged tracks. 

With all the above considerations in mind, 
we concentrate on the lighter {\it stau} (${\stau}_1$)
to be the NLSP with lifetime large enough to penetrate collider
detectors like the muons themselves. Using the spectrum generator
of the package ISAJET 7.69 \cite{isajet}, we find that a large
mSUGRA parameter space can realize this scenario of a right-sneutrino
LSP and stau NLSP, provided that $m_0 < m_{1/2}$ and one has 
$tan\beta$ of the order of 10 and above, the latter condition being
responsible for a larger left-right off-diagonal term in the 
stau mass matrix (and thus one smaller eigenvalue). 

 It is worth mentioning that the 
region of the mSUGRA parameter space where we work 
is consistent with all experimental bounds, including both collider 
and low-energy constraints (such as the LEP and Tevatron constraints 
on the masses of Higgs, gluinos, charginos and so on as well as those from 
$b \rightarrow s\gamma$, correction to 
the $\rho$-parameter, ($g_\mu - 2$) etc.). Our choice of parameters in the
$m_0-m_{1/2}$ plane would correspond to a stau-LSP without the
right-sneutrino in the (super)particle spectrum, which is ruled out in
the context of $\chi^0_1$-LSP scenario \cite{djouadi}.
However with the right-sneutrino as the
LSP, we find this choice to be a preferable and well-motivated option.
The $\snu_1$-LSP arising out of such a choice becomes a viable constituent
of cold dark matter, though not necessarily the only one.

We mostly focus on the region of the
parameter space where $m_{{\stau}_1} > m_{{\tilde \nu}_1} + m_W$ 
is satisfied, so the dominant decay mode is the 
two-body decay of the NLSP, ${\stau}_1 \to \tilde{\nu}_1  W$. In
Figure~\ref{decay} we show the lifetime of such long-lived stau's 
plotted against
the universal gaugino mass parameter $m_{1/2}$ for a particular choice
of~$(m_0,A,sign(\mu)~{\rm and}~\tan\beta)$. 
One must however note that having too large a
value for $m_{1/2}$ will make the gluinos and squarks too heavy to be
produced at colliders. We have chosen the case of
degenerate neutrinos, since this permits the largest possible
values of the Yukawa couplings. The fact that the lifetimes are
so large even with this choice is a convincing demonstration
of the long-lived nature of the stau-NLSP, as far as collider detectors
are concerned. In Table~\ref{param.tbl} we present two benchmark 
points for our study of such long-lived staus at the LHC.
In the next section we discuss the signatures of the stau NLSP at the
LHC, concentrating on the two types of final states mentioned in the
introduction. Using the formulae given by Moroi {\it et al.} in reference
4, the contribution to the cold dark matter relic density is found to be about
one order of magnitude below the acceptable value of $\Omega h^2$.
While this leaves room for additional sources of dark matter, the scenario
presented here is consistent from the viewpoint of over-closure of the universe.

\section{Signatures of stau-NLSP at LHC}
In this section we discuss the signatures of the long-lived 
stau-NLSP at the LHC and concentrate on two different final states,
viz.
\begin{itemize}
\item $2 \stau_1 + 2({\rm or~more})~jets~(p_T > 100~{\rm GeV})$
\item $2 \stau_1 + {\rm dimuon} + 2({\rm or~more})~jets~
(p_T > 100~{\rm GeV})$
\end{itemize} 
Keeping the above signals in mind, we focus on the two benchmark points
listed in Table~\ref{param.tbl} and study their signatures at the LHC,
for an integrated luminosity of $30~fb^{-1}$. 
\bt
\begin{center}
\begin{tabular}{|c|c|c|}
\hline\hline
{\bf Parameter}& {\bf Benchmark point 1}&{\bf Benchmark point
2}\\\hline
            &$m_0=100~GeV,~m_{1/2}=600~GeV$
&$m_0=110~GeV,~m_{1/2}=700~GeV$\\
mSUGRA input&$A=100~GeV,~sgn(\mu)=+$&$A=100~GeV,~sgn(\mu)=+$\\
            &$\tan\beta=30$ &$\tan\beta=10$\\\hline\hline
$|\mu|$                              &694&810\\\hline
$\mel,\mml$                          &420&486\\
$\mer,\mmr$                          &251&289\\
$m_{\snu_{eL}},m_{\snu_{\mu L}}$     &412&479\\
$m_{\snu_{\tau L}}$                  &403&478\\
$m_{\snu_{iR}}$                      &100&110\\
$m_{\stau_1}$                        &187&281\\
$m_{\stau_2}$                        &422&486\\\hline\hline
$m_{\chi^0_1}$                       &243&285\\
$m_{\chi^0_2}$                       &469&551\\
$m_{\chi^0_3}$                       &700&815\\
$m_{\chi^0_4}$                       &713&829\\
$m_{\chi^{\pm}_1}$                   &470&552\\
$m_{\chi^{\pm}_2}$                   &713&829\\\hline
$m_{\tilde{g}}$                      &1366&1574\\\hline\hline
$m_{\tilde{u}_L},m_{\tilde{c}_L}$    &1237&1424\\
$m_{\tilde{u}_R},m_{\tilde{c}_R}$    &1193&1373\\
$m_{\tilde{d}_L},m_{\tilde{s}_L}$    &1239&1426\\
$m_{\tilde{d}_R},m_{\tilde{s}_R}$    &1189&1367\\
$m_{\tilde{t}_1}$                    &984&1137\\
$m_{\tilde{t}_2}$                    &1176&1365\\
$m_{\tilde{b}_1}$                    &1123&1330\\
$m_{\tilde{b}_2}$                    &1161&1358\\\hline\hline
$m_{h^0}$                            & 118&118\\
$m_{H^0}$                            & 712&941\\
$m_{A^0}$                            & 707&935\\
$m_{H^{\pm}}$                        & 717&944\\\hline\hline
\end{tabular}
\caption {\small
\it{Proposed benchmark points for study of stau-NLSP scenario
in the SUGRA fold with right-sneutrino LSP. All superparticle masses
are given in GeV. Due to very small mixing, $\snu_1  \simeq \snu_R$. 
The top mass, $m_t~=~171.4$ GeV \cite{mtop} has been used for running 
the parameters.}}
\label{param.tbl}
\end{center}
\et

The low energy mass spectrum schematically takes the following form:
$$m_{\tilde{\nu}_1}<m_{\tilde{\tau}_1}<m_{\tilde{\chi}_1^0}<
m_{\tilde{e}_1,\tilde{\mu}_1}<...<m_{\tilde{g}}$$
which would suggest that all superparticle productions at the LHC would
finally end at the sneutrino LSP. However, as discussed in the previous
section and also evident from Figure~\ref{decay}, 
the lifetime of the $\stau_1$-NLSP is quite large, 
ranging between a few seconds to a few minutes, making it stable in the 
context of collider studies. So it will almost always decay outside 
the detector, leaving characteristic signals like charged tracks, 
with large transverse momenta. In fact this would be quite a contrast to the
traditionally thought of SUSY signals with {\it large missing transverse energy.} 
This fact acts as the main thrust of our work, as the stable stau 
will behave just like a muon. However, in the absence of spin identification,
these staus will behave more like very heavy muon-like particles 
with $\beta(=v/c)< 1$ and such heavy charged particles will
have high specific ionisation due to their slow motion within the
detector. Studies exist in the literature which have looked at such heavy
and stable charged particles and considered their ionisation properties
and time of flight as distinctive qualities which separate them from the
muons \cite{longlived}. 
In this work we take a qualitatively different approach towards
studying the long-lived staus which would be based more on an analysis
pertaining to studying the kinematics of processes producing such
particles at the LHC. We will see through our analysis, 
that by simply looking at certain kinematic distributions, one can 
distinguish such long-lived staus from the muons. In the kinematic
plots presented later in this section, we have neglected the energy
loss of the staus reaching the muon chambers. Including this loss
is matter of detailed detector analysis, and are unlikely to
change the general features of the tracks, which are qualitatively
rather distinctive.
 
It should be noted that small variations of the values of parameters
used here may lead to one selectron and smuon state each to be also
lighter than the lightest neutralino. In that case, although the
only two-body decay available to each of them lead to the lighter
sneutrino states of the first two families, such decays are
tremendously suppressed due the same reason as in the case
of the stau-NLSP. On the other hand, each of them can
have three-body decay, mediated, for example, by a virtual 
neutralino, into a stau, a tau and an electron(muon). This 
channel will prevent the selectrons(smuons) from being long-lived,
while they will give rise to additional events with  
stau charged tracks.

We now take up the above mentioned signals one by one and present our
results in the following subsections. 
\subsection {{$2 \stau_1$ + hard jets}}
For our simulation we have generated 10 million events for SUSY
particle production, and allowed their decay through all possible 
modes in to lighter stau. The signals mentioned here arise mostly from
the direct decay of gluinos and squarks, produced via strong interaction, 
into the lightest neutralino,
with the latter decaying into a tau and a lighter stau, and
the tau decaying hadronically in turn. Cascades through other
neutralinos and charginos supplement the rates to a moderate extent.
We have used PYTHIA 6.409 \cite{pythia} for our event generation
and interfaced it with ISASUGRA, the SUSY-spectrum generator routine
contained in ISAJET 7.69. Following the procedure outlined in the previous
section, we have introduced the right-sneutrino 
in the ISAJET program to generate the mSUGRA spectrum with a
right-sneutrino LSP with its corresponding renormalization
group equation RGE (eqn. 5) also included. 
To make the $\chi^0_1$ decay in PYTHIA we have modified the decay table 
to include its decay to $\stau_1 \bar{\tau}$ and demand that the 
$\chi^0_1$ is not the LSP.

The parton densities have been evaluated at $Q=2m_{\tilde{\tau}_1}$ 
using CTEQ5L \cite{cteq}, 
and the renormalization scale and factorization scale are
$$\mu_F=Q=\mu_R$$
(On setting the scale at the squark/gluino masses the rates
are reduced by about 20\%, and none of the general conclusions is
altered). The effects of both initial state radiation (ISR) and final 
state radiation (FSR) have also been taken into account. We have included 
the effects of hadronization and multiple interaction with the
help of PYTHIA hadronization schemes.

To define jets we use the simple-minded jet cone algorithm implemented
in PYTHIA through the subroutine PYCELL. The basic parameters we have used 
in this cluster-finding routine are as follows: 
\begin{itemize}
\item The jet conical width is $\Delta R_{jj}\ge0.7$,  
      where $$\Delta R_{jj}=\sqrt{\Delta \eta_{jj}+\Delta \phi_{jj}}$$
    where  $\Delta \eta_{jj}$ and $\Delta \phi_{jj}$ are intervals in
    rapidity and azimuthal angle.

\item The Gaussian smearing of the $E_\perp$ with a width proportional
to $1.2 E_\perp$ is used to take into account the energy resolution of
the detector.

\item The summed $E_\perp$ of the jet (consisting of all cells within
the cone of radius $R$ in the ($\eta,\phi$) plane) should be greater 
than 50 GeV to be accepted as a jet.

\item The $\eta$ coverage range for jets is taken to be $|\eta|\leq 3.0$. 
\end{itemize}
To select our final states, we demand the following requirements 
(called basic cuts) on our sample events: 
\begin{itemize}
\item Each $\stau_1$ should have $p_T > 30$ GeV.
\item Both the $\stau_1$'s should satisfy $|\eta|\leq 2.5$, to ensure
that they lie within the coverage of the muon detector.
\item  $\Delta R_{\stau_1\stau_1} \geq 0.2$, to ensure that the $\stau_1$'s 
are well resolved in space.
\item  At least two jets with $p_T > 100$ GeV (hard jets).
\item In addition we have rejected events having photons with 
$|\eta_{\gamma}|\leq 2.5$ and ${p_T}_{\gamma}\ge 25$ GeV.
\end{itemize}
\begin{figure}[ht]
\begin{center}
\epsfig{file=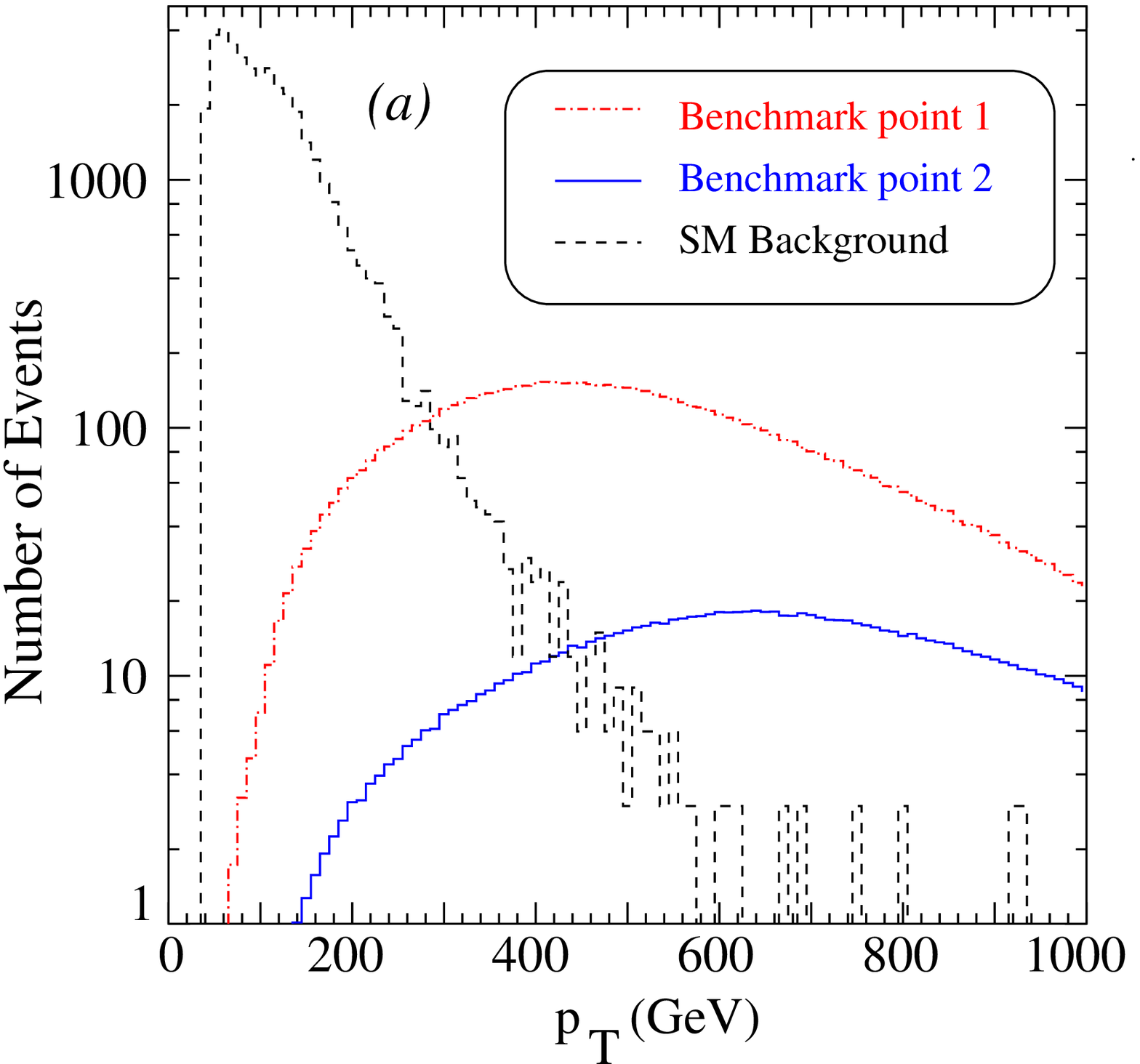,width=2.6in,height=2.6in}
\epsfig{file=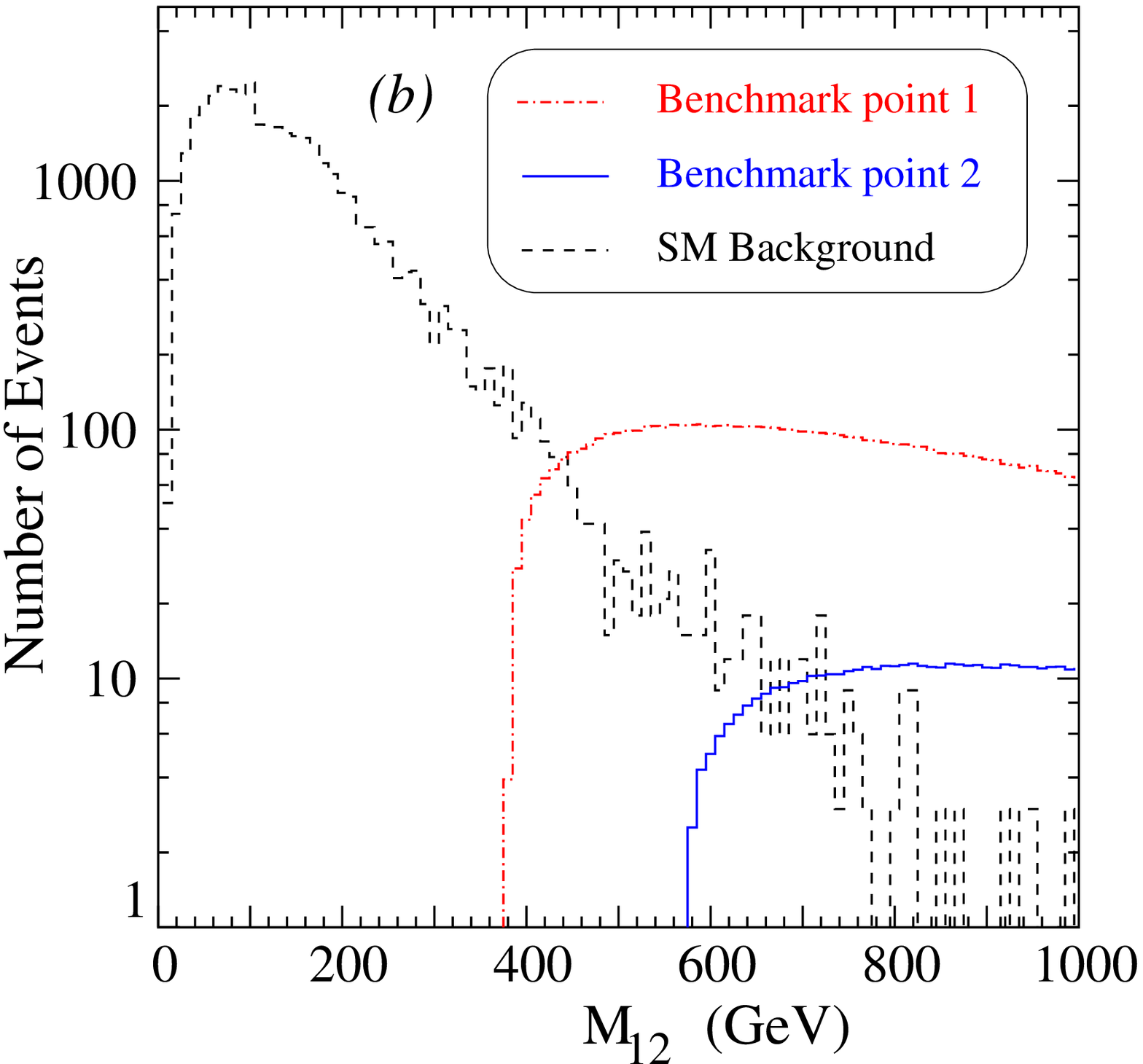,width=2.6in,height=2.6in}
\caption{\small    
\it{Kinematic distributions for the signal
$2\stau_1 + (\ge 2)$ hard jets. In (a) the transverse momentum 
distributions for the harder $\stau_1$ is shown and (b) shows the 
invariant mass distribution for the $\stau_1$ pair. The dash-dot-dash
(red) histograms are for benchmark point 1 and the solid (blue)
histogram for benchmark point 2. The dashed (black) histograms show the 
corresponding SM background.}}
\label{s1nc}
\end{center}
\end{figure}
As the charged tracks pass through the muon chamber, it is probable 
that the muonic events will fake our signal. Therefore, events with
\begin{figure}[htb]
\begin{center}
\epsfig{file=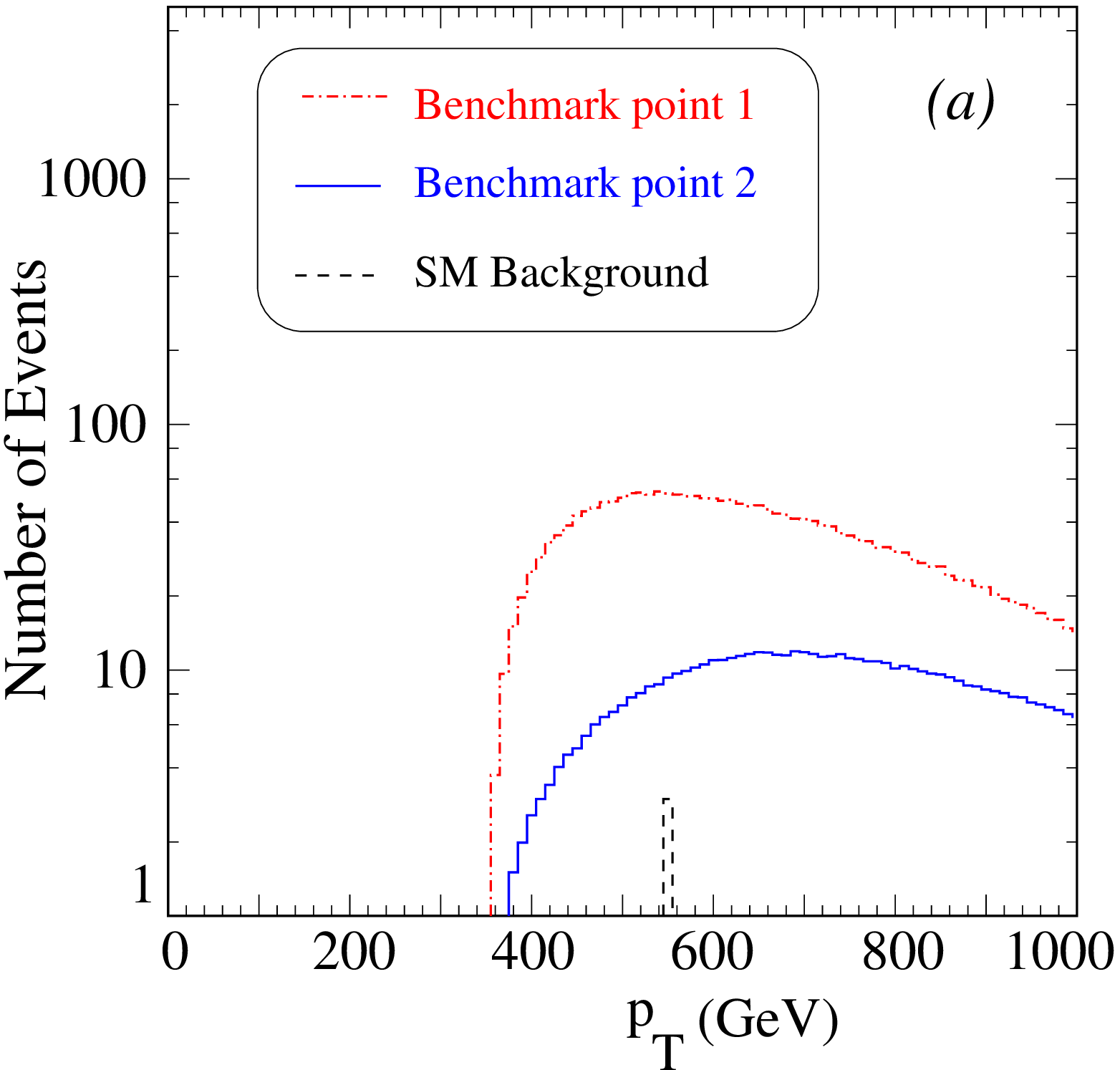,width=2.6in,height=2.6in}
\epsfig{file=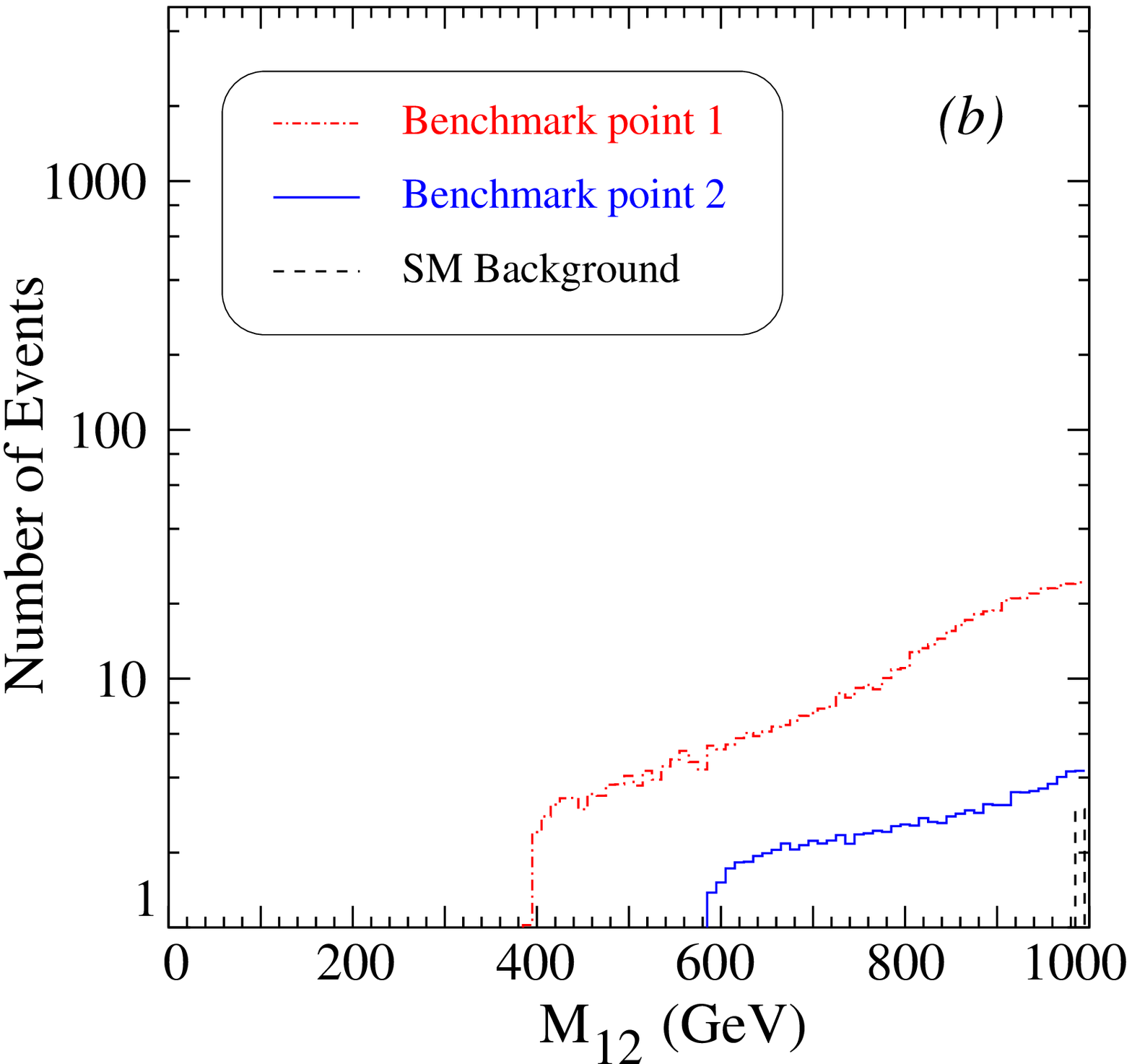,width=2.6in,height=2.6in}
\caption{\small
\it{Kinematic distributions for the signal
$2\stau_1 + (\ge 2)$ hard jets after imposing the stronger cut,
$p_T > 350$ GeV on both charged tracks. In (a) the transverse momentum
distributions for the harder $\stau_1$ is shown and (b) shows the
invariant mass distribution for the $\stau_1$ pair. 
We follow the same notation as in Figure \ref{s1nc}.}}
\label{s1wc}
\end{center}
\end{figure}
two or more hard jets and two central muons will {\it prima facie}    
constitute our standard model background. The leading contribution
to such final states satisfying our basic cuts comes
from top-pair production and its subsequent decay into 
dimuons, with similar topology as that of the signal. The 
sub-leading contributions consist in weak boson pair production. 
For performing the background analysis we again used the same criteria as 
stated above for the signal, with all kinematic features of
the long-lived charged tracks attributed to the muons. 

We would like to emphasize here that the demand for two (at least) 
hard jets with $p_T\ge100$ GeV may itself be expected to 
act as a `killer' to background, as we are well aware of the fact 
that events with large number of high $p_T$ jets is itself a `hint' to 
SUSY provided that we do not have other candidates for physics beyond 
the SM. However, SUSY signals have this advantage ordinarily through
the occurrence of large missing-$\not{p_T}$. In this case, the hard 
stau tracks assume the role of the latter. 
The other important point to note is that the background is almost 
completely reducible with the imposition of stronger event selection
criteria, as we shall see below.
In Figure~\ref{s1nc},
we present distributions of a few observables where we could distinguish the 
signal from backgrounds. These are the transverse momentum ($p_T$) of the 
harder charged track (and the harder muon in the case of backgrounds) 
and the invariant mass ($M_{12}$) of the two charged tracks (or dimuons). 
In addition, the radii of curvatures of the stau tracks will also lie
in a clearly distinguishable range, as seen from the $p_T$-distributions.

\bt[hb]   
\begin{center}
\begin{tabular}{|c|c|c|c|}
\hline
{\bf{Cuts}}& {\bf{Background}}&{\bf{Benchmark point 1}}&{\bf{Benchmark
point 2}}\\\hline  
{Basic}& 39617&8337&1278\\\hline
{Basic $+~p_T>350$ GeV}& 5&2587&737\\\hline
\end{tabular}  
\caption{\small 
\it{The expected number of events for the signal
and background with the different cuts imposed on the selection of
events. We have assumed an integrated luminosity of $30~fb^{-1}$ at the
LHC to generate the events.}}
\label{s1evt.tbl}
\end{center}
\et
Figure~\ref{s1nc} indicates that the basic cuts retain a large number of
background events. A substantial fraction of these events come from the
Z-peak, as evident from the invariant mass distributions. However,
on examining the  $p_T$-distributions,
we find it most convenient to eliminate them by imposing a stronger
$p_T$ cut on {\em both the charged tracks}. This cut has been
set at 350 GeV. The effect is rather dramatic, and makes the signal
stand out clearly for both the benchmark points, as can be seen from 
Table~\ref{s1evt.tbl}.

\begin{figure}[htb]
\begin{center}
\epsfig{file=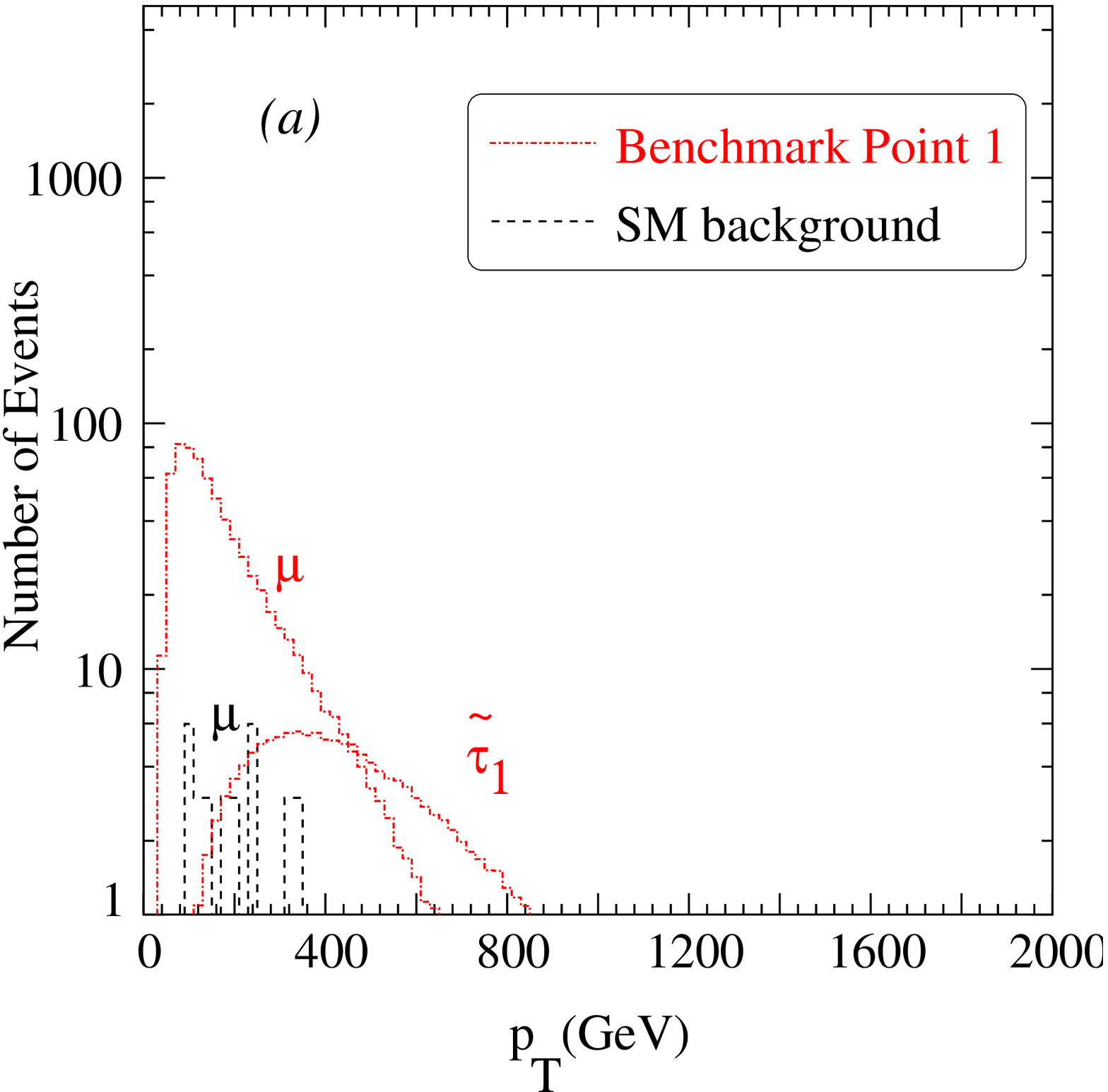,width=2.5in,height=2.5in}
\epsfig{file=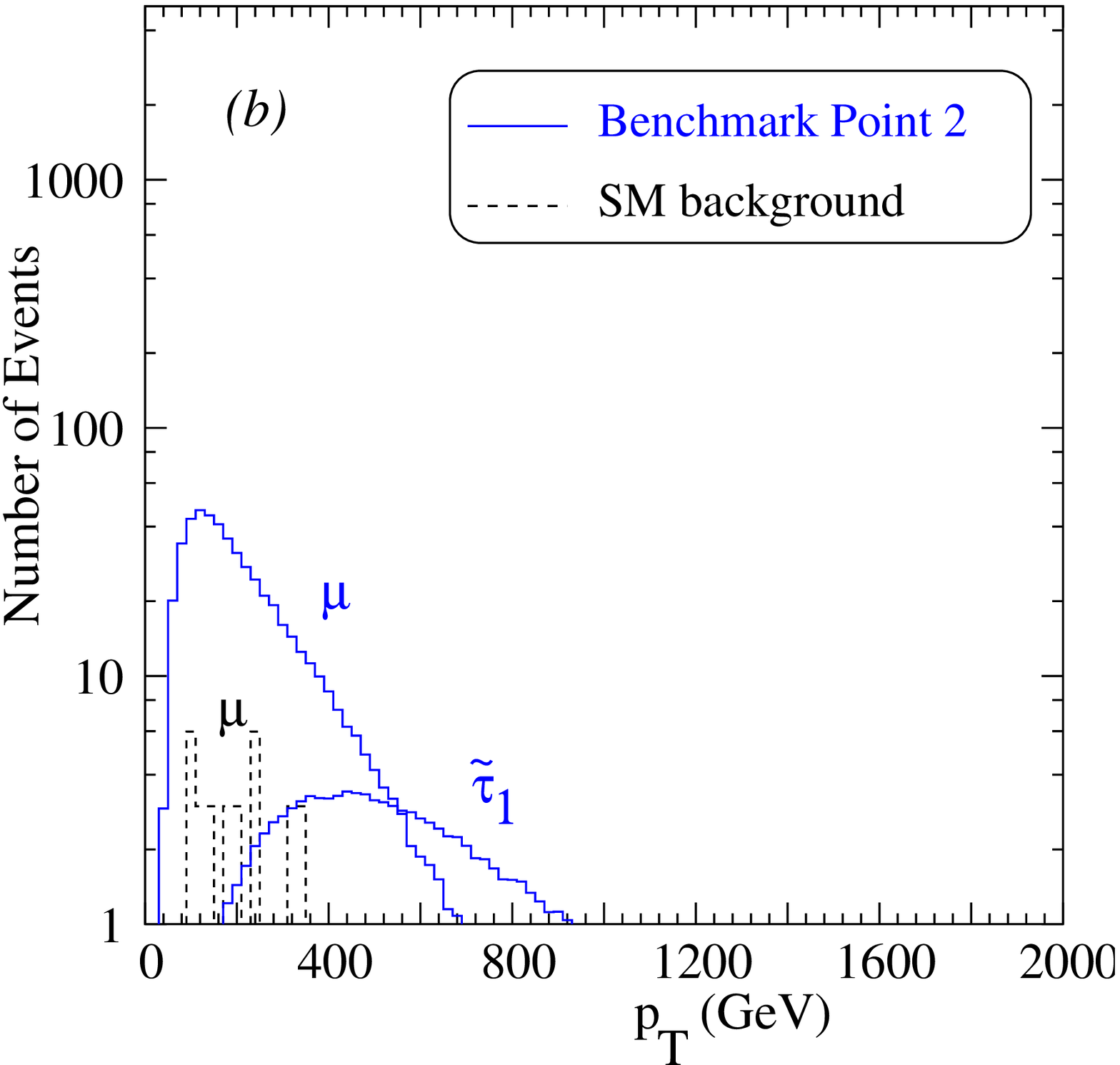,width=2.5in,height=2.5in}
\epsfig{file=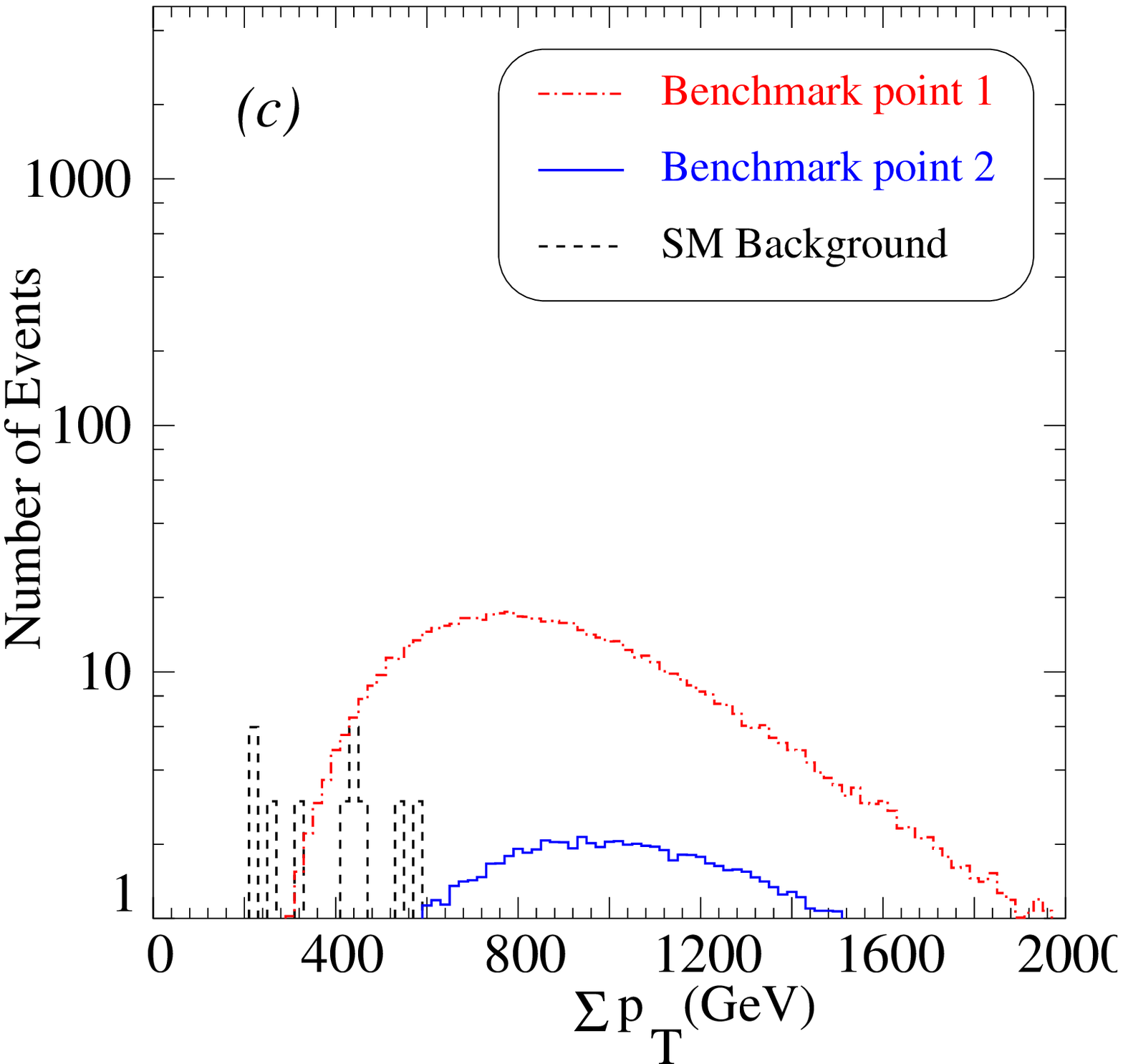,width=2.5in,height=2.5in}
\caption{\small
\it{Kinematic distributions for the signal
$2\stau_1 + dimuon + (\ge 2)$ hard jets with basic cuts.
In (a) and (b) the transverse momentum
distributions for the harder $\stau_1$ and harder muon of the signal 
is shown and (c{\small
\it{Kinematic distributions for the signal
$2\stau_1 + dimuon + (\ge 2)$ hard jets with basic cuts.
In (a) and (b) the transverse momentum
distributions for the harder $\stau_1$ and harder muon of the signal 
is shown and (c) shows the scalar sum of $p_T$'s of (dimuon +
2$\stau_1$) of the signal. We follow the same notation as in Figure 
\ref{s1nc}.}}) shows the scalar sum of $p_T$'s of (dimuon +
2$\stau_1$) of the signal. We follow the same notation as in Figure 
\ref{s1nc}.}}
\label{s2wc}
\end{center}
\end{figure}

Figure~\ref{s1wc} shows the corresponding distributions after the 
backgrounds are
nearly gone. These constitute the real `signal distributions', and can be
used to extract information, for example, about the mass
of the stau-NLSP and other SUSY parameters.

Finally, one has to remember that, in addition to the removal of 
SM backgrounds, one needs to distinguish the signal mentioned above from 
that of a SUSY scenario where the lightest neutralino is the LSP. There, 
too, a pair of high-$p_T$ muons can arise from cascades, together with 
hard jets. However, as has already been mentioned, the neutralino
LSP causes the missing-$p_T$ distribution to be much harder, at least
when the LSP mass is related through universality conditions to
the higher charginos and neutralinos which, in turn, are constrained by 
results from the Large Electron Positron (LEP) collider. Consequently,
signals in the scenario under investigation here will be more
severely affected when a missing-$p_T$ cut is imposed. 
To take a specific example, the signal rate at benchmark point 1
becomes about 48\% of its original value when the requirement of
a minimum missing-$p_T$ of 100 GeV is imposed. In contrast, at a 
a nearby point in the parameter space 
($m_0~=~200$ GeV, $m_{1/2}~=~600$ GeV, $A~=~100$ GeV, $\tan\beta~=~30$,
$sgn(\mu)$ = +ve), the same missing-$p_T$ cut allows about 97\%
survival of the $jets~+~dimuons~+\not{p_T}$ signal. Thus the response
to missing-$p_T$ cuts turns out to be an effective tool of differentiation
between our signal and that coming from MSSM, at least when R-parity 
is conserved.

\subsection {\bf{Dimuon and two staus with two or more jets}}
With the stringent demand on the hardness of the jets, this is a very
clean signal, albeit less copious than the previous one. Such final states
will require cascade decays of gluinos and squarks involving 
the charginos and heavier neutralinos. The same
`basic cuts' are imposed here, too, which are found to be sufficient 
in drastically reducing the SM backgrounds in the form of four muons
together with two or more jets with $p_T > 100$ GeV.
\footnote{It should be noted that there are some uncertainties in 
the way PYTHIA models jet production, beyond two-jet final states.
However, we retained PYTHIA results here, partly because the main point
we wish to illustrate is unaffected, and partly due to the want of
a completely reliable jet-production algorithm including SUSY.} 
 
In Figure~\ref{s2wc} we plot the corresponding distributions for the above
final states.
It is the requirement of such energetic jets that removes the
backgrounds to a large extent, as can be seen from Table~\ref{s2evt.tbl}. 
Furthermore, the plot of the scalar sum of the individual transverse
momenta of the charged tracks in the muon chamber, in
Figure~\ref{s2wc}(c), indicates that 
a cut of 600 GeV on this sum washes out the backgrounds completely.
The advantage of applying this cut is that there is no loss
of signal events in the case of a relatively heavy NLSP.
In Figure~\ref{s2wc}(a) and (b), the $p_T$-distributions of the 
(harder) muon and the corresponding stau-track are seen to have a 
substantial overlap. Therefore, a distinction between them based on
the thickness of the tracks as well as the information provided
by measurement of the `time-of-flight' can be useful here.
\bt[hb]
\begin{center}
\begin{tabular}{|c|c|c|c|}
\hline
{\bf{Final States}}& {\bf{Background}}&{\bf{Benchmark point
1}}&{\bf{Benchmark point 2}}\\\hline
$2\stau_1 + 2\mu$&83&689&103\\\hline
$2\stau_1 + 2\mu +(\ge 2)$ hard jets& 29&686&103\\\hline
$2\stau_1 + 2\mu +(\ge 2)$ hard jets& 0 &553&89 \\
 ($\sum p_T > 600$ GeV)             &   &   &    \\\hline
\end{tabular}
\caption{\small
\it{The expected number of events for the signal
and background with the different cuts imposed on the selection of
events. $\sum p_T$ corresponds to the scalar sum of the individual
transverse momenta of the charged tracks in the muon chamber. 
Choice of integrated luminosity is the same as
Table~\ref{s1evt.tbl}.}}
\label{s2evt.tbl}
\end{center}
\et
\section {\bf{Summary and conclusions}}
We have studied a SUSY scenario with a stau-NLSP and an
overwhelmingly right-chiral sneutrino as the LSP, where
the sneutrino is at least partially responsible for the
cold dark matter of the universe. A mass spectrum corresponding
to such a scenario can be motivated in a SUGRA framework,
including right-chiral sneutrinos whose masses remain practically
frozen at the universal scalar mass at the SUSY breaking scale.
We find that the superparticle
cascades culminating into the production of stau-pairs give
rise to very distinct signals of such a scenario. Although the
charged tracks of the quasi-stable staus tend to fake
muonic signals in the muon chamber, our analysis reveals
considerable difference in their kinematic characters.
Such difference can be used in a straightforward way 
to distinguish between the long-lived staus and the muons,
and also to eliminate all standard model backgrounds. Since
the mass spectrum under consideration here is as probable
as one with a neutralino LSP in mSUGRA, further study of
all possible ways of uncovering its signature at the LHC
should be of paramount importance. 

\bigskip
\noindent
{\bf {Acknowledgments}} We thank S. Mrenna for useful
technical advice. We are also grateful to S. Banerjee, S. Bhattacharya, 
A. Das, S. P. Das, A. K. Datta and 
P. Skands for helpful discussions. This work was partially supported
by the Department of Atomic Energy, Government of India, through
a project funded under the Xth 5-year Plan.

\vskip 5pt


\begin{thebibliography}{99}
\bibitem{reviews1}
  For reviews see, for example, 
  H.~P.~Nilles,
  Phys.\ Rept.\  {\bf 110}, 1 (1984);
  H.~E.~Haber and G.~L.~Kane,
  Phys.\ Rept.\  {\bf 117}, 75 (1985);
  M.~Drees,
  arXiv:hep-ph/9611409;
  S.~P.~Martin,
  arXiv:hep-ph/9709356, and references therein;
  D.~J.~H.~Chung, L.~L.~Everett, G.~L.~Kane, S.~F.~King, J.~D.~Lykken
  and L.~T.~Wang,
  Phys.\ Rept.\  {\bf 407}, 1 (2005)
  [arXiv:hep-ph/0312378].
\bibitem{reviews2}
  For reviews see, for example, 
  S.~Dawson, E.~Eichten and C.~Quigg,
  Phys.\ Rev.\ D {\bf 31}, 1581 (1985);
  H.~Baer, C.~h.~Chen, F.~Paige and X.~Tata,
  Phys.\ Rev.\ D {\bf 52}, 2746 (1995)
  [arXiv:hep-ph/9503271];
  H.~Baer, C.~h.~Chen, F.~Paige and X.~Tata,
  Phys.\ Rev.\ D {\bf 53}, 6241 (1996)
  [arXiv:hep-ph/9512383].
\bibitem{revnu}
For a review, see for example,
  R.~N.~Mohapatra {\it et al.},
  arXiv:hep-ph/0510213 and references therein.
\bibitem{rneut}
  A.~T.~Alan and S.~Sultansoy,
  J.\ Phys.\ G {\bf 30}, 937 (2004)
  [arXiv:hep-ph/0307143];
  T.~Asaka, K.~Ishiwata and T.~Moroi,
  Phys.\ Rev.\ D {\bf 73}, 051301 (2006)
  [arXiv:hep-ph/0512118].
\bibitem{hebbeker}
  T.~Hebbeker,
  Phys.\ Lett.\ B {\bf 470}, 259 (1999)
  [arXiv:hep-ph/9910326].
\bibitem{arkani}
  N.~Arkani-Hamed, L.~J.~Hall, H.~Murayama, D.~R.~Smith and N.~Weiner,
  Phys.\ Rev.\ D {\bf 64}, 115011 (2001)
  [arXiv:hep-ph/0006312];
  D.~Hooper, J.~March-Russell and S.~M.~West,
  Phys.\ Lett.\ B {\bf 605}, 228 (2005)
  [arXiv:hep-ph/0410114].
\bibitem{gopala}
  C.~L.~Chou and M.~E.~Peskin,
  Phys.\ Rev.\ D {\bf 61}, 055004 (2000)
  [arXiv:hep-ph/9909536];
  A.~de Gouvea, S.~Gopalakrishna and W.~Porod,
  JHEP {\bf 0611}, 050 (2006)
  [arXiv:hep-ph/0606296].
\bibitem{gravitinoLSP}
  J.~L.~Feng, A.~Rajaraman and F.~Takayama,
  Phys.\ Rev.\ Lett.\  {\bf 91}, 011302 (2003)
  [arXiv:hep-ph/0302215];
  J.~L.~Feng, A.~Rajaraman and F.~Takayama,
  Phys.\ Rev.\ D {\bf 68}, 063504 (2003)
  [arXiv:hep-ph/0306024];
  J.~R.~Ellis, K.~A.~Olive, Y.~Santoso and V.~C.~Spanos,
  Phys.\ Lett.\ B {\bf 588}, 7 (2004)
  [arXiv:hep-ph/0312262];
  J.~L.~Feng, S.~Su and F.~Takayama,
  Phys.\ Rev.\ D {\bf 70}, 075019 (2004)
  [arXiv:hep-ph/0404231];
  A.~Ibarra and S.~Roy,
  arXiv:hep-ph/0606116.
\bibitem{gmsbnlsp}
  D.~A.~Dicus, B.~Dutta and S.~Nandi,
  Phys.\ Rev.\ Lett.\  {\bf 78}, 3055 (1997)
  [arXiv:hep-ph/9701341];
  S.~Ambrosanio, G.~D.~Kribs and S.~P.~Martin,
  Phys.\ Rev.\ D {\bf 56}, 1761 (1997)
  [arXiv:hep-ph/9703211];
  D.~A.~Dicus, B.~Dutta and S.~Nandi,
  Phys.\ Rev.\ D {\bf 56}, 5748 (1997)
  [arXiv:hep-ph/9704225];
  K.~M.~Cheung, D.~A.~Dicus, B.~Dutta and S.~Nandi,
  Phys.\ Rev.\ D {\bf 58}, 015008 (1998)
  [arXiv:hep-ph/9711216];
  J.~L.~Feng and T.~Moroi,
  Phys.\ Rev.\ D {\bf 58}, 035001 (1998)
  [arXiv:hep-ph/9712499];
  P.~G.~Mercadante, J.~K.~Mizukoshi and H.~Yamamoto,
  Phys.\ Rev.\ D {\bf 64}, 015005 (2001)
  [arXiv:hep-ph/0010067].
\bibitem{coanni}
  S.~Ambrosanio, G.~D.~Kribs and S.~P.~Martin,
  Phys.\ Rev.\ D {\bf 56}, 1761 (1997)
  [arXiv:hep-ph/9703211];
  A.~V.~Gladyshev, D.~I.~Kazakov and M.~G.~Paucar,
  Mod.\ Phys.\ Lett.\ A {\bf 20}, 3085 (2005)
  [arXiv:hep-ph/0509168];
  T.~Jittoh, J.~Sato, T.~Shimomura and M.~Yamanaka,
  Phys.\ Rev.\ D {\bf 73}, 055009 (2006)
  [arXiv:hep-ph/0512197].
\bibitem{Moroi2}
  T.~Asaka, K.~Ishiwata and T.~Moroi,
  arXiv:hep-ph/0612211.
\bibitem{longlived}
  M.~Drees and X.~Tata,
  Phys.\ Lett.\ B {\bf 252}, 695 (1990);
  A.~Nisati, S.~Petrarca and G.~Salvini,
  Mod.\ Phys.\ Lett.\ A {\bf 12}, 2213 (1997)
  [arXiv:hep-ph/9707376];
  S.~P.~Martin and J.~D.~Wells,
  Phys.\ Rev.\ D {\bf 59}, 035008 (1999)
  [arXiv:hep-ph/9805289];
  J.~L.~Feng and T.~Moroi,
  Phys.\ Rev.\ D {\bf 61}, 095004 (2000)
  [arXiv:hep-ph/9907319];
  S.~Ambrosanio, B.~Mele, S.~Petrarca, G.~Polesello and A.~Rimoldi,
  JHEP {\bf 0101}, 014 (2001)
  [arXiv:hep-ph/0010081];
  W.~Buchmuller, K.~Hamaguchi, M.~Ratz and T.~Yanagida,
  Phys.\ Lett.\ B {\bf 588}, 90 (2004)
  [arXiv:hep-ph/0402179];
  J.~R.~Ellis, A.~R.~Raklev and O.~K.~Oye,
  JHEP {\bf 0610}, 061 (2006)
  [arXiv:hep-ph/0607261].
\bibitem{stoppers}
  K.~Hamaguchi, Y.~Kuno, T.~Nakaya and M.~M.~Nojiri,
  Phys.\ Rev.\ D {\bf 70}, 115007 (2004)
  [arXiv:hep-ph/0409248];
  J.~L.~Feng and B.~T.~Smith,
  Phys.\ Rev.\ D {\bf 71}, 015004 (2005)
  [Erratum-ibid.\ D {\bf 71}, 0109904 (2005)]
  [arXiv:hep-ph/0409278].
  K.~Hamaguchi, M.~M.~Nojiri and A.~de Roeck,
  arXiv:hep-ph/0612060.
\bibitem{neuexp}
  S.~Fukuda {\it et al.}  [Super-Kamiokande Collaboration],
  Phys.\ Lett.\ B {\bf 539}, 179 (2002)
  [arXiv:hep-ex/0205075];
\bibitem{Araki:2004mb}
  T.~Araki {\it et al.}  [KamLAND Collaboration],
  Phys.\ Rev.\ Lett.\  {\bf 94}, 081801 (2005)
  [arXiv:hep-ex/0406035];
  E.~Aliu {\it et al.}  [K2K Collaboration],
  Phys.\ Rev.\ Lett.\  {\bf 94}, 081802 (2005)
  [arXiv:hep-ex/0411038];
  Y.~Ashie {\it et al.}  [Super-Kamiokande Collaboration],
  Phys.\ Rev.\ D {\bf 71}, 112005 (2005)
  [arXiv:hep-ex/0501064].
\bibitem{expbound}
  W.~M.~Yao {\it et al.}  [Particle Data Group],
  J.\ Phys.\ G {\bf 33}, 1 (2006);
  E.~Barberio {\it et al.}  [Heavy Flavor Averaging Group (HFAG)],
  arXiv:hep-ex/0603003.
\bibitem{wmap}
  D.~N.~Spergel {\it et al.},
  arXiv:astro-ph/0603449.
\bibitem{isajet}
  F.~E.~Paige, S.~D.~Protopopescu, H.~Baer and X.~Tata,
  arXiv:hep-ph/0312045.
\bibitem{djouadi}
  A.~Djouadi, M.~Drees and J.~L.~Kneur,
  JHEP {\bf 0603}, 033 (2006)
  [arXiv:hep-ph/0602001].
\bibitem{mtop}   
  E.~Brubaker {\it et al.}  [Tevatron Electroweak Working Group],
  arXiv:hep-ex/0608032.
\bibitem{pythia} 
  T.~Sjostrand, S.~Mrenna and P.~Skands,
  JHEP {\bf 0605}, 026 (2006)
  [arXiv:hep-ph/0603175].
\bibitem{cteq} 
  H.~L.~Lai {\it et al.},
  Phys.\ Rev.\ D {\bf 51}, 4763 (1995)
  [arXiv:hep-ph/9410404].
\end{thebibliography}
\end{document}